\documentclass[12pt]{iopart}
\usepackage[dvips]{graphicx,color}
\begin{document}

\title[Rigorous theory of nuclear fusion rates in a
  plasma]{Rigorous theory of nuclear fusion rates in a plasma}

\author{Lowell S. Brown, David C. Dooling, and Dean L. Preston}

\address{Los Alamos National Laboratory, Los Alamos, New Mexico 87545, USA}

\eads{\mailto{brownl@lanl.gov}, \mailto{dcd@lanl.gov}, 
         \mailto{dean@lanl.gov}}

\begin{abstract}

Real-time thermal field theory is used to reveal the
structure of plasma corrections to nuclear reactions.  Previous
results are recovered in a fashion that clarifies their nature,
and new extensions are made.  Brown and Yaffe  have
introduced the methods of effective quantum field theory into
plasma physics They are used here to treat the interesting
limiting case of dilute but very highly charged particles
reacting in a dilute, one-component plasma. The highly charged
particles are very strongly coupled to this background plasma.
The effective field theory proves that this mean field solution
plus the one-loop term dominate; higher loop corrections are
negligible even though the problem involves strong coupling.
Such analytic results for very strong coupling are rarely
available, and they can serve as benchmarks for testing computer
models.

\end{abstract}

\pacs{{\bf 24.10.-\bf i}, {\bf 52.25.-\bf b}}

\section{General Formulation}

 A nuclear reaction, which we schematically indicate by
  $ 1 + 2 \to 3 + 4 $,
takes place over a very short distance in comparison with particle
separations in a plasma. Hence, it can be described by an
effective local Hamiltonian density
 \begin{equation}
 {\cal H}({\bf x},t) = g \, {\cal K}({\bf x},t) 
                    + g \, {\cal K}^\dagger({\bf x},t) \,.
\end{equation}
 The operator ${\cal K}$  describes, with interaction strength
$g$, the destruction of the initial particles and the creation of the 
 final particles; the operator ${\cal K}^\dagger$  does the reverse.
 Fermi's golden rule presents the rate as
\begin{eqnarray}
\Gamma &=& \int_{-\infty}^{+\infty} dt \, e^{iQt/\hbar} \, 
\int (d^3{\bf x}) \left\langle {\cal K}^\dagger({\bf x},t) {\cal K}(0) 
\right\rangle_\beta \,.
\label{genresult}
\end{eqnarray}
The angular brackets $ \langle \cdots \rangle_\beta$  denote the
thermal average; $Q$ is the reaction energy release.

The extension of imaginary time thermodynamic theory to
include real time behavior was initiated long ago by
Schwinger \cite{Sch} and Keldysh \cite{Kel}.
Using this method as a basis, a detailed analysis \cite{BDP} 
shows that when the particles entering into the nuclear reaction 
can be treated by Maxwell-Boltzmann statistics,
\begin{eqnarray}
\Gamma
&=& g^2 \, {n^{(0)}_1 \, n^{(0)}_2 \over \lambda_1^{-3} \, \lambda_2^{-3}}
 \,\int_{-\infty}^{+\infty} dt
\, e^{i Q t/ \hbar} \, \int (d^3{\bf x}) \, \hat{Z}_{C}
\left[ {\cal V} {\hbar \over i} {\delta \over \delta \phi} \right]
\nonumber\\
&& \qquad
\left.
\langle {\bf 0} , {\bf 0} , -i \beta\hbar |
       {\bf x}, {\bf x} , t\rangle^{V_C \, \phi}_{1+2} \,\,
\langle {\bf x}, {\bf x} , t |
 {\bf 0} ,{\bf 0},0 \rangle^{V_C \, \phi}_{3+4}\right|_{\phi=0} \,,
\label{allah}
\end{eqnarray}
 with the functional integral definition
\begin{eqnarray}
\hat{Z}_{C}[\phi] & = &  Z^{-1} \, 
\int \prod_b \left[ d\psi^*_b d\psi_b \right] \,
\exp\left\{ {i \over \hbar} \int_{C} d s \, L
               \right\}
\nonumber\\
&& \qquad\qquad
\, \exp\left\{ {i \over \hbar} \int_{C} d s
\int (d^3{\bf y}) \rho({\bf y}, s)
           \phi({\bf y}, s) \right\} \,.
\label{ballah}
\end{eqnarray}

\begin{figure}
\begin{picture}(280,280)(-100,0)
\put(50,260){\line(1,0){150}}
\put(50,280){\line(0,-1){120}}
\thicklines \put(50,265){\vector(1,0){50}}
\put(100,265){\line(0,-1){10}}
\put(100,255){\vector(-1,0){50}}
\put(49.9,255){\vector(0,-1){75}}
\thinlines
\put(105,245){$t$}
\put(70,272){$\bf{C_{+}}$}
\put(70,240){$\bf{C'_{-}}$}
\put(30,180){-i$\beta$}
\put(20,210){$\bf{{C''}_{-}}$}
\end{picture}
\vspace{-2.3 in}
\caption{
{The ${\bf C_+}$
portion represents the interactions
between the plasma and the final reaction particles 
that appear in
$
\langle {\bf x}, {\bf x} , t |
 {\bf 0} ,{\bf 0},0 \rangle^{V_C \, \phi}_{3+4}
$.
The $\bf{C_-}$ part is needed for the plasma interactions with the
initial reaction particles that enter into
$
\langle {\bf 0} , {\bf 0} , -i \beta\hbar |
       {\bf x}, {\bf x} , t\rangle^{V_C \, \phi}_{1+2}
$.
This contour has the real $\bf{C'_-}$ and purely imaginary
$\bf{{C''}_-}$} parts.}
\label{contour}
\end{figure}
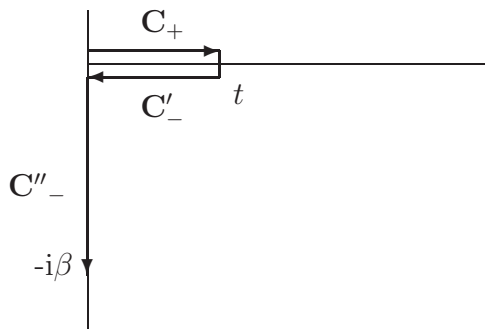

\noindent  All the field variables $\psi$ in the plasma
Lagrangian $L$ and plasma charge density $\rho$ are functions
of the spatial coordinate ${\bf y}$, and the generalized time
variable $s$ runs along the contour  $ {\bf C}$ shown  in Fig.~1.
The reacting particles have thermal wave lengths $\lambda_{1,2}$
and, with no plasma interactions, they would have
number densities $n^{(0)}_{1,2}$.

The structure of the result (\ref{allah}) is easy to understand.  
The two transformation  functions
$ \langle \cdots | \cdots \rangle^{V_C \, \phi} $  describe the
propagation of the initial and final  particles
that undergo the nuclear reaction.  The $V_C$  superscripts indicate
that these particles interact  via their mutual Coulomb forces.  The
$\phi$  superscripts indicate that these particles also interact with
an arbitrary external potential. The operator
$
\hat{Z}_{C}
\left[ {\cal V} {\hbar \over i} {\delta \over \delta \phi} \right]
$
produces the Coulomb interactions
between the reacting particles and the background  plasma.

In essentially all cases of interest, one can neglect the real time
portions $\bf{C_+}$  and $\bf{C_-'}$  because
 of the factor $\exp\{i Q t / \hbar \}$:  the relevant real time
scale is $\hbar / Q$,  a time very much shorter than any characteristic
plasma time.
In many cases of interest, $ \kappa \, r_{\rm max} \ll 1$,
 where $\kappa$  is the Debye wave number and 
$r_{\rm max}$  is the turning point radius of the Coulomb interaction
between the initial particles.  Then the rate reduces to 
\cite{DGC,BS,BDP}
\begin{equation}
\Gamma = \Gamma_C \, { N^{(0)}_1 \over N_1} \,
{ N^{(0)}_2 \over N_2 } \,  
 { N_{1+2} \over N^{(0)}_{1+2} } \,\,.
\label{xrated}
\end{equation}
Here $\Gamma_C$  is the nuclear reaction rate for a thermal,
Maxwell-Boltzmann distribution of the initial (1,2) particles at
temperature $T$  but with no plasma background.  The rate 
$ \Gamma_C$  does contain the full effects of the Coulomb forces
between the reacting particles.  The number $N_a^{(0)}$  is the
particle number obtained for a free gas grand canonical ensemble 
with chemical potential $\mu_a$.  The number $N_a$  is the particle
number of this species $a$ with the same chemical potential  $\mu_a$ 
but now interacting in the plasma. The subscripts $1+2$ denote a
composite particle of charge $(Z_1+Z_2)e$. 

\section{Method Illustrated By Improving The Ion Sphere Model}

 The simplest example has a weakly interacting one-component plasma,
$
g \ll 1
$,
where
$
g = \beta  e^2 \kappa / 4\pi \,.
$
The effective field theory of Brown and Yaffe \cite{BY} shows
that 
\begin{equation}
N_p = { N_p^{(0)} \over{\cal Z} } \, \int [d\chi] \, e^{-S[\chi]} \,,
\label{clever}
\end{equation}
where
\begin{equation}
\fl \qquad\quad
S[\chi] =
 \int (d^3{\bf r}) \left[
{\beta \over 2}  \Big( \nabla \chi({\bf r})  \Big)^2
\!\! - n  \Bigg( e^{ie\beta  \chi({\bf r})} - 1 
\! - ie \beta  \chi({\bf r}) \Bigg) \, 
\! - i Z_p e \beta \, \delta({\bf r}) \chi({\bf r}) \right] . 
\label{act}
\end{equation}
The normalizing partition
function ${\cal  Z}$  is defined by the functional
integral whose action omits the $\delta$ function term in 
Eq.~(\ref{act}).  The tree approximation is given by 
$S[i \phi_{\rm cl}({\bf r})]$ 
with
\begin{equation}
- \nabla^2 \phi_{\rm cl} ({\bf r}) = e n \left[ 
  e^{-\beta e \phi_{\rm cl}({\bf r})} - 1 \right]
+ Z_p e \, \delta({\bf r}) \,.
\label{obey}
\end{equation}
 This is the familiar Debye-Huckle form, but now
placed in a systematic
perturbative expansion where error can be
ascertained. 
Including the one-loop correction gives
\begin{equation}
N_p = N_p^{(0)} \, 
{ {\rm Det}^{1/2} \left[ - \nabla^2 + \kappa^2 \right] \over
{\rm  Det}^{1/2} \left[ - \nabla^2 + \kappa^2 \, 
    e^{-\beta e \, \phi_{\rm cl}} \right] } \,
\exp\left\{- S[i \phi_{\rm cl}] \right\} \,.
\label{stuff}
\end{equation}

We work in the limit where $Z_p$  is so large that 
$gZ_p \gg 1$.  The point charge $Z_pe/4\pi r$  part of
$\phi_{\rm cl}({\bf r})$  is large and dominates over a large range.
This validates the Salpeter ion sphere model which approximates
$
\Big[ 1 - \exp\{ - \beta e \phi_{\rm cl}({\bf r}) \} \Big]
\simeq \theta\left( r_0 - r \right) \,.
$
The total plasma charge in this uniform sphere must cancel the 
impurity charge, and so
$
r_0^3 = {3 gZ_p / \kappa^3 } \,.
$
The first correction to the leading Salpeter solution can also
be computed in analytic form  except for a numerical integral.
Including this correction gives, with $ {\cal C} = 0.8498 \cdots $, 
\begin{equation}
 - S[i\phi_{\rm  cl}] + Z_p \simeq 
 { 3 Z_p \over 10 } \, \left( 3gZ_p \right)^{2/3} \,
    \left\{ 1 + { 10 \, {\cal C} \over \left( 3 g Z_p ) \right) }
   \right\} \,. 
\label{next}
\end{equation} 
\begin{figure}
\centerline{
\includegraphics[height=5cm]{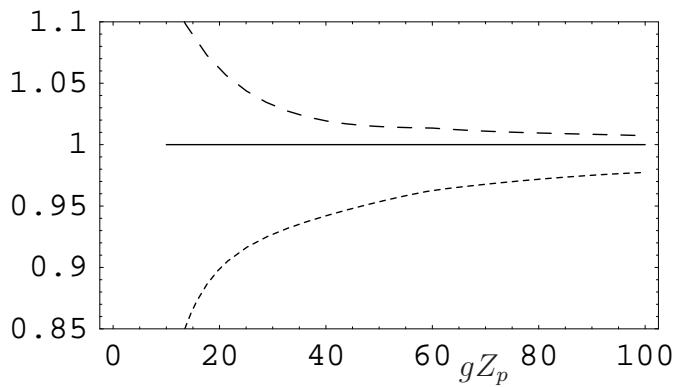}
\hspace{-3.25cm} {$gZ_p$} 
}
\caption{Ratios of $S[i\phi_{\rm cl}] - Z_p$ for
the ion sphere model result [short-dashed line]
and the corrected ion sphere model [long-dashed line]
to the exact numerical action.}
\label{ratio}
\end{figure}

Brown and Yaffe \cite{BY} have shown that the one-loop correction
for the background plasma with no impurity ions present is given by
\begin{equation}
{\rm Det}^{-1/2} \left[ - \nabla^2 + \kappa^2 \right] 
= \exp\left\{ \int (d^3{\bf r}) \, { \kappa^3 \over 12 \pi} \right\} \,.
\end{equation}
In our limit the term
$
\kappa^2 \, \exp\left\{ - \beta e \phi({\bf r}) \right\} 
$
in the one-loop determinant can be treated as being very slowly 
varying except when it appears in a final volume integral. Thus, 
\begin{eqnarray}
{ {\rm Det}^{1/2} \left[ - \nabla^2 + \kappa^2 \right] \over 
{\rm Det}^{1/2} \left[ - \nabla^2 + \kappa^2 \,
      e^{- \beta e \phi_{\rm cl} } \right]  }
&=&
\exp\left\{-  { \kappa^3 \over 12 \pi} \, {4\pi \over 3} \, 
r_0^3 \right\} 
 = \exp\left\{ - {1\over3} \, g Z_p \right\} \,.
\label{oneloop}
\end{eqnarray} 
This result is physically obvious.  The ion of high
$Z_p$ carves out a hole of radius $r_0$ in the original plasma.
The original plasma is unchanged outside this hole.  Corrections
smooth out the sharp boundaries and produce only higher-order
terms.  The original plasma had a vanishing electrostatic
potential everywhere, and the potential in the ion sphere picture
now vanishes outside the sphere of radius $r_0$.  Thus the
thermodynamic potential of the plasma is reduced by the amount
that was originally contained within the sphere of radius $r_0$,
and this is exactly what is stated to one-loop order in
Eq.(\ref{oneloop}).  This argument carries on to the higher loop
terms as well. A term involving $n$ loops carries a factor $g^n$.
The presence of the impurity modifies this to be $ Z g^n$. With
$g$ sufficiently small, all the higher-order loops make
negligible contributions.  The corrected impurity number $N_p$ is
hence given by Eq's.~(\ref{oneloop}) and (\ref{next}) inserted
into Eq.~(\ref{stuff}).

The number relation expresses the nuclear rate (\ref{xrated}) 
in terms of the tree contribution.  
Including the first correction to the ion sphere
result gives 
\begin{eqnarray}
\Gamma &=& \Gamma_C \, \exp\left\{ {3\over10} \, (3 g)^{2/3} \,
   \left[ \left( Z_1 + Z_2 \right)^{5/3} - Z_1^{5/3} - Z_2^{5/3}
         \right] \right\} 
\nonumber\\
&& \qquad\qquad \times \,
\exp\left\{ \left( { \,9\, \over g} \right)^{1/3} \, {\cal C} \,
      \left[ \left( Z_1 + Z_2 \right)^{2/3} - 
   Z_1^{2/3} - Z_2^{2/3} \right]
   \right\} \,.
\end{eqnarray}
The first line agrees with the calculation of Salpeter
\cite{Sal}; the second is a new correction.

The number correction for the number of impurity
ions $N_p$  placed in the weakly coupled background plasma with
number $N$  can be used to construct the grand canonical partition
function ${\cal Z}$  for the combined system by integrating the generic
relation
$
N = \partial \ln {\cal Z} /  \partial \beta \mu \,. 
$
To simply bring out the main point, we now include only the 
leading terms. Standard thermodynamic relations then lead to the 
equation of state 
\begin{eqnarray}
p V &=& \left\{ N - Z_p \, {(3gZ_p)^{2/3} \over  10} \,
                 \, N_p \right\} \, T \,.
\end{eqnarray}
Although $ N_p / N$  may be small, there is a large 
pressure modification if $Z_p$  is large.


\section*{References}

\begin{thebibliography}{99}

\bibitem{Sch}
        Schwinger J
        1961 {\it Journ. of Math. Phys.} {\bf 2} 407

\bibitem{Kel}
        Keldysh L V
        {1964 \it Zh. Eksp. Teor. Fiz.} {\bf 47} 1515 
        [1965 {\it Sov. Phys. JETP} {\bf 20} 1018]

\bibitem{BDP}
        Brown L S, Dooling D C and Preston D L
	in preparation

\bibitem{DGC}
        DeWitt H E, Graboske H C and Cooper M S
        1973 {\it Astrophys. J.} {\bf 181} 439 

\bibitem{BS}
        Brown L S and Sawyer R F
        1997 {\it Rev. Mod. Phys.} {\bf 69} 411

\bibitem{BY}
        Brown L S and Yaffe L G 
        2001 {\it Phys. Rep.} {\bf 340} 1

\bibitem{Sal}
        Salpeter E E 
        1954 {\it Aust. J. Phys.} {\bf 7} 373 

\end{thebibliography}
\end{document}